\documentclass[sigconf]{acmart}
\AtBeginDocument{%
  }

\copyrightyear{2026}
\acmYear{2026}
\setcopyright{cc}
\setcctype{by}
\acmConference[CHI EA '26]{Extended Abstracts of the 2026 CHI Conference on Human Factors in Computing Systems}{April 13--17, 2026}{Barcelona, Spain}
\acmBooktitle{Extended Abstracts of the 2026 CHI Conference on Human Factors in Computing Systems (CHI EA '26), April 13--17, 2026, Barcelona, Spain}
\acmDOI{10.1145/3772363.3799094}
\acmISBN{979-8-4007-2281-3/2026/04}

\usepackage{xspace}  
\usepackage{multirow}

\begin{document}


\title{iPoster: Content-Aware Layout Generation for Interactive Poster Design via Graph-Enhanced Diffusion Models}
\author{Xudong Zhou}
\orcid{0009-0004-9106-6773}
\affiliation{%
  \institution{Beijing Institute of Technology}
  \city{Beijing}
  \country{China}
}
\email{3120230947@bit.edu.cn}

\author{Jinyuan Liang}
\affiliation{%
  \institution{Beijing Institute of Technology}
  \city{Beijing}
  \country{China}
}
\email{3220241667@bit.edu.cn}

\author{Qiuyi Guo}
\affiliation{%
  \institution{Beijing Institute of Technology}
  \city{Beijing}
  \country{China}
}
\email{3120250950@bit.edu.cn}

\author{Guozheng Li}
\authornote{Guozheng Li is the corresponding author.}
\affiliation{%
  \institution{Beijing Institute of Technology}
  \city{Beijing}
  \country{China}
}
\email{guozheng.li@bit.edu.cn}

\renewcommand{\shortauthors}{Zhou et al.}
\renewcommand{\shorttitle}{iPoster}

\begin{abstract}
We present iPoster, an interactive layout generation framework that empowers users to guide content-aware poster layout design by specifying flexible constraints.
iPoster enables users to specify partial intentions within the intention module, such as element categories, sizes, positions, or coarse initial drafts. 
Then, the generation module instantly generates refined, context-sensitive layouts that faithfully respect these constraints. 
iPoster employs a unified graph-enhanced diffusion architecture that supports various design tasks under user-specified constraints.
These constraints are enforced through masking strategies that precisely preserve user input at every denoising step.
A cross content-aware attention module aligns generated elements with salient regions of the canvas, ensuring visual coherence. 
Extensive experiments show that iPoster not only achieves state-of-the-art layout quality, but offers a responsive and controllable framework for poster layout design with constraints.
\end{abstract}



\maketitle

\section{Introduction}
\label{sec: Introduction}
Content-aware layout generation aims to arrange appealing visual elements, such as texts, logos, underlays, within a predefined canvas.
This method has been widely applied to the design of e-commerce posters~\cite{guo2021vinci, o2015designscape}, magazine covers~\cite{jahanian2013recommendation, yang2016automatic}, and various multimedia presentation scenes.

In recent years, numerous studies~\cite{zheng2019content, cglgan, posterlayout, icvt, radm, layoutdit, lin2023layoutprompter, hsu2025postero} have been proposed to address the problems of content-aware layout generation.
These methods have demonstrated competitive performance in generating aesthetically pleasing layout results. 
Nevertheless, several limitations persist. 
First, current methods do not support direct user interaction, making it challenging to explicitly integrate user’ preferences and expectations, such as the category, position, and size of visual elements, into the layout generation process.
Second, the generated layouts frequently exhibit undesirable issues, such as occlusion of essential canvas regions, overlap, and misalignment of layout elements.

To address these challenges, we introduce iPoster, an interactive layout generation framework that integrates diffusion models and graph neural networks (GNNs)~\cite{scarselli2008graph}. 
The interaction module enables users to specify heterogeneous constraints, and the generation module conditions the diffusion process on these constraints to produce layouts that faithfully reflect user intent. 
At its core, the generation model features a two-stage GNN-based cross content-aware attention module: the first stage processes a fully connected graph of layout elements and salient regions to capture spatial balance, while the second stage models topological relationships between elements and image patches to extract alignment features—enabling coherent, content-aware layouts with reduced overlap and preserved visual quality.

\begin{figure*}[hbtp]
\centering
\includegraphics[width=1.\linewidth]{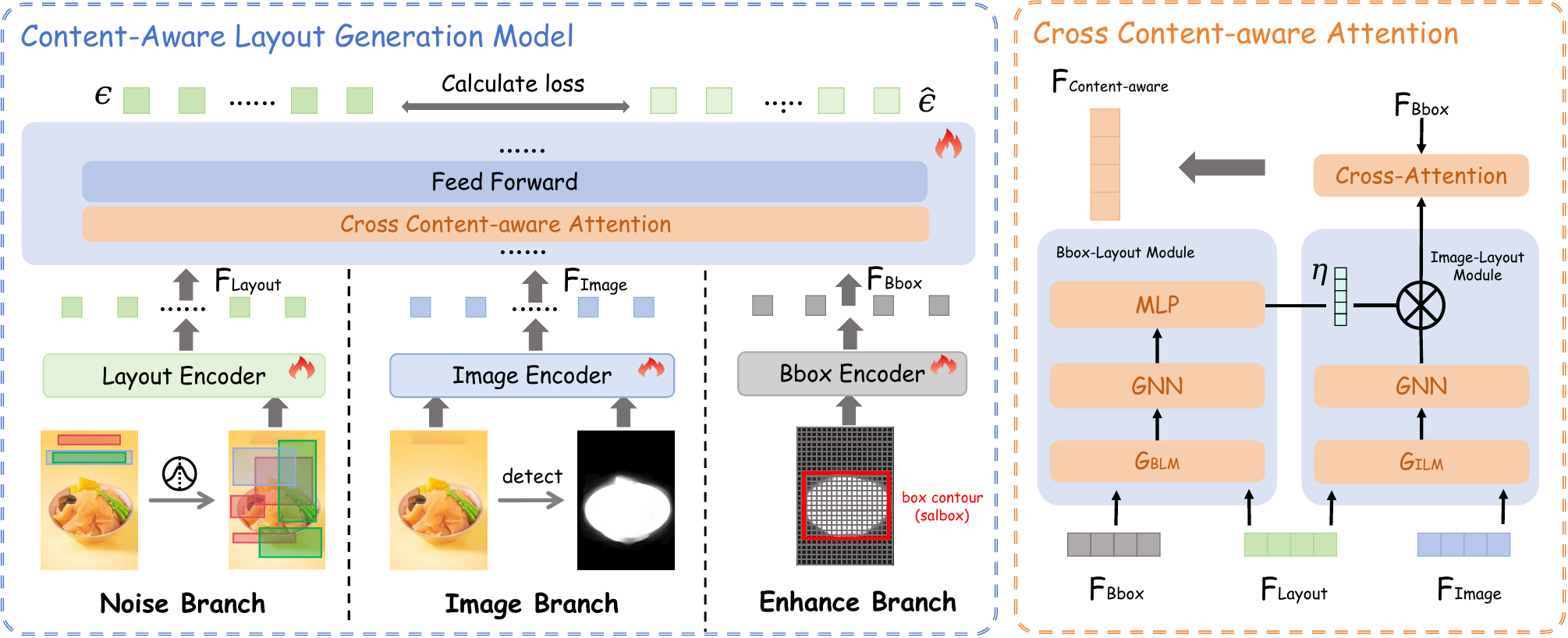}
\caption{
The overall training framework of content-aware layout generation model (shown in the left).
The core module is Cross Content-aware Attention Module (shown in the right).
}
\label{fig:pipe}
\end{figure*}

To validate the effectiveness of our method, we conducted extensive experiments in various constrained tasks using two public datasets: PKU~\cite{posterlayout} and CGL~\cite{cglgan}. Quantitative and qualitative results demonstrate that our method effectively addresses the issues of occlusion, overlap, and misalignment. 
Our key contributions are summarized as follows.
\begin{itemize}

    \item A user-oriented interaction framework to support flexible input constraints, enabling users to intuitively specify various design intents during the generation process, thereby meeting users' diverse expectations.

    \item A GNNs-based content-aware layout generation model to explicitly model the spatial and topological relationships between layout elements and image patches, achieving improved spatial balance among layout elements and canvas regions. 
\end{itemize}

\section{Related Work}

\subsection{Content-aware Layout Generation}

ContentGAN~\cite{zheng2019content} first used image semantics to guide layout design. CGL-GAN~\cite{cglgan} further used saliency maps to improve subject identification. To capture the structural dependencies among layout elements, DS-GAN~\cite{posterlayout} modeled sequential dependencies in layout with a CNN-LSTM architecture, while ICVT~\cite{icvt} used a conditional VAE to predict layout elements autoregressively. Moving beyond GAN-based approaches, RADM~\cite{radm} leveraged a diffusion framework for better alignment between visual and textual elements. Complementing these modeling advances, RALF~\cite{ralf} addressed the issue of limited training data with a retrieval-augmented approach. LayoutDiT~\cite{layoutdit} introduced a transformer-based diffusion model for layout distribution modeling.

In addition, several methods have explored the use of large language models (LLMs) for layout generation~\cite{posterllava, patnaik2025aesthetiq, shi2025layoutcot, hsu2025postero}. 
Although they exhibit robust generalization, they typically incur inference latency, impose considerable hardware and compute requirements, and are prone to hallucination errors characteristic of LLMs, thereby limiting their scalability in real‑world deployments.

\subsection{Interactive Poster Design}
Previous interactive graphic design methods~\cite{shi2020emog, zhao2020iconate, hegemann2023computational, o2015designscape, guo2021vinci} have demonstrated promising potential to foster user creativity through real-time assistance. Lee et al.~\cite{lee2010designing} introduced an interactive example gallery to support users in the design process. For poster design, DesignScape~\cite{o2015designscape} provides context-aware interactive suggestions for non-expert users, and Vinci~\cite{guo2021vinci} synthesizes poster layouts from user-provided text and hierarchical layer structures.

However, many interaction-based methods suffer from a key limitation: while they emphasize user-driven design through manual placement or rule-based adjustments, they lack an engine capable of automatically generating aesthetically plausible layouts.
In contrast, while the content-aware layout generation methods reviewed in Section 2.1 excel at generating visually plausible layouts, they operate entirely automatically, leaving users little room to inject intent, correct errors, or provide real-time guidance for the design process.
This contradiction creates a gap between user-expressive control and intelligent automation.

\section{Method}

\begin{figure*}[htbp]
\centering
\includegraphics[width=1.\linewidth]{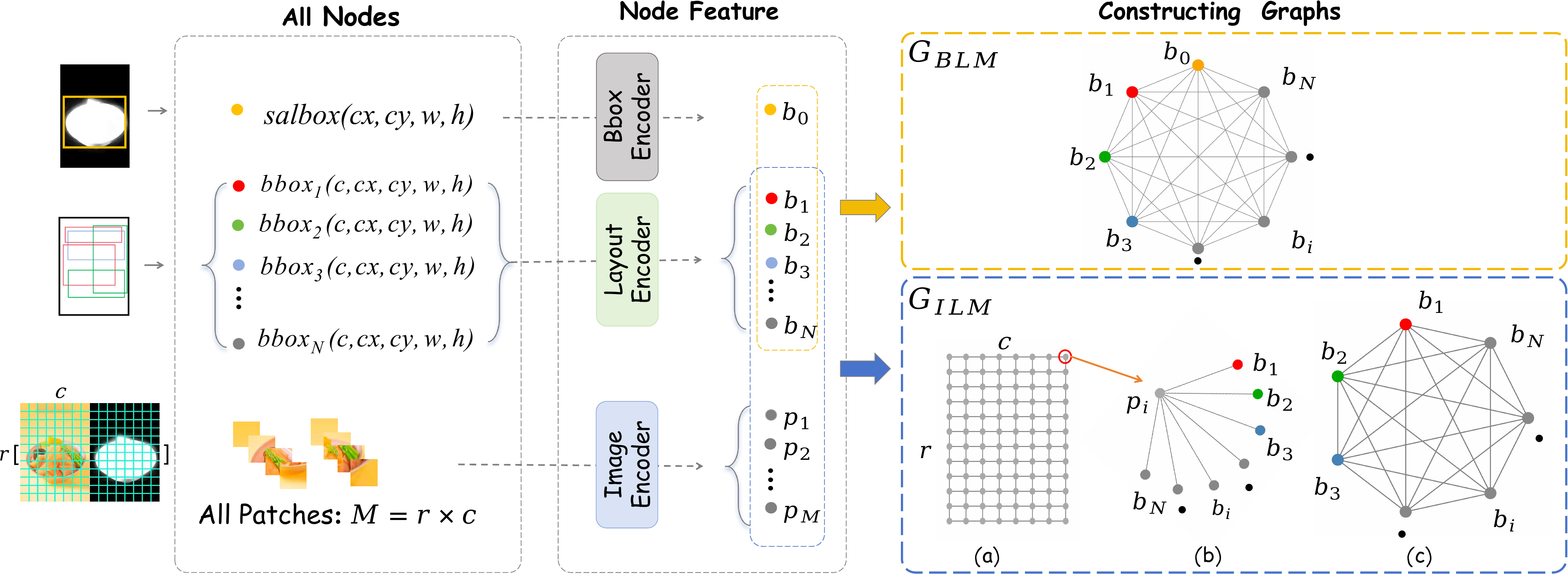}
\caption{
The construction process of $G_{BLM}$ and $G_{ILM}$. 
For each $bbox_i$, $c$ is the element category, $cx$ and $cy$ are its center coordinates, and $w$ and $h$ are its width and height. 
Each image is divided into $r$ rows and $c$ columns of patches.
See Section~\ref{sec:Constructing Graphs} for details.
}

\label{fig:graphs}
\end{figure*}

\subsection{Preliminaries: User Constraints}
\label{sec:Preliminaries}
Informed by prior studies on graphic design workflows and constraint-based layout systems~\cite{kong2022blt, jiang2023layoutformer}, we analyze common patterns in user-specified requirements during poster design. This analysis reveals four recurrent constraint paradigms, which we formalize to guide content-aware layout generation.

$Category \rightarrow Size + Position (C \rightarrow S + P)$.
In this task, users can specify the category of each layout element according to their design intent, and the model automatically infers a plausible size and position for every element.
This task is suitable for scenarios where users explicitly specify the categories of the layout elements to be placed. 
For example, given a product poster, the user may intend to add a logo, two text (e.g., a promotional description and a discounted price), and an underlay.
The specific positions and sizes of these elements are adaptively generated by the model.

$Category + Size \rightarrow Position (C + S \rightarrow   P)$.
Here, users are allowed to define both the category and size of each layout element, while the model generates spatially coherent positions that respect the specified constraints.
For example, users can assign a reasonable bounding box size to a logo and, for each text element, determine the bounding box dimensions based on its character count and the desired font size.

$Completion$.
In this task, users can partially author a layout by fixing the category, size, and position of a subset of elements; the model then completes the layout by generating compatible configurations for the remaining elements, conditioned on the user-provided anchors.
For example, a user may wish to place the product logo in a specific location on the poster to emphasize the brand identity of the product.

$Refinement$. This task aims to optimize suboptimal layouts. Users provide a coarse initial layout, and the model iteratively optimizes it into a polished and well-structured layout in a few steps.
For example, when the initially generated layout does not meet user expectations, this task can be used for iterative refinement and improvement.

\subsection{Content-Aware Layout Generation Model}

\subsubsection{Model Inputs}
\label{sec:Model Inputs}

Figure~\ref{fig:pipe} shows the design details of our content-aware generation model.

\textbf{Image and Saliency Map.}
The input to the image encoder is a four-channel image constructed by merging the canvas $I$ and the saliency map $S$. 
Details on obtaining Canvas $I$ and the saliency map $S$ can be found in the supplementary materials.
It is passed through a ViT-based Image Encoder to obtain $F_{Image}$.

\textbf{Layout.}
The parameters of layout elements are represented as $L = \{(c_i, cx_i, cy_i, w_i, h_i )\}$, where $c_i$ is the category label and $[cx_i, cy_i, w_i, h_i]$ is the normalized bounding box.  
The Layout Encoder encodes $L$ to get $F_{Layout}$.

\textbf{Saliency Bounding Box.}
Following the LayoutDiT~\cite{layoutdit}, we extract the bounding box of the salient region from the saliency map by thresholding the pixel values. This bounding box, denoted as $salbox= \{cx, cy, w, h\}$, preserves the spatial information of the main area on the canvas. 
$salbox$ is passed through the Bbox Encoder to get $F_{Bbox}$.

\subsubsection{Constructing Graphs}
\label{sec:Constructing Graphs}
We construct two distinct graphs for the Bbox-Layout Module (BLM) and the Image-Layout Module (ILM), denoted as $G_{BLM}$ and $G_{ILM}$.
Their construction process is shown in Figure~\ref{fig:graphs}.

\textbf{Construction of $G_{BLM}$}.
As shown in Figure~\ref{fig:graphs}, 
$G_{BLM}$ is a fully connected graph consisting of N+1 nodes, including one $salbox$ node and the $N$ layout element nodes.

\textbf{Construction of $G_{ILM}$}.
$G_{ILM}$ consists of the $M$ image patch nodes and the $N$ layout element nodes, yielding a total of $M+N$ nodes.
Edges in this graph are added in the following three steps: 
($a$) The $M$ image patch nodes are connected based on spatial adjacency, forming a grid graph. 
($b$) Each image patch node $p_i$ is connected to all $N$ layout element nodes $(b_1, b_2, ...b_N)$. 
($c$) The $N$ layout element nodes are connected to each other.

\subsubsection{Cross Content-aware Attention Module}\label{sec:Content-aware Module}

For content-aware layout generation, it is crucial to balance two aspects: the spatial relationship between layout elements and the visual areas of canvas and the internal arrangement among the layout elements. 
The former ensures that important visual areas are not occluded by design elements, while the latter minimizes issues such as overlapping  between layout elements.
To achieve this, we design a cross content-aware attention module that dynamically adjusts the relationship between layout elements and canvas content. 
This module consists of the BLM and ILM (both GNN-based), and the Cross-Attention Module, as shown in the right part of Figure~\ref{fig:pipe}.

\begin{figure*}[hbtp]
\centering 
\includegraphics[width=1.\linewidth]{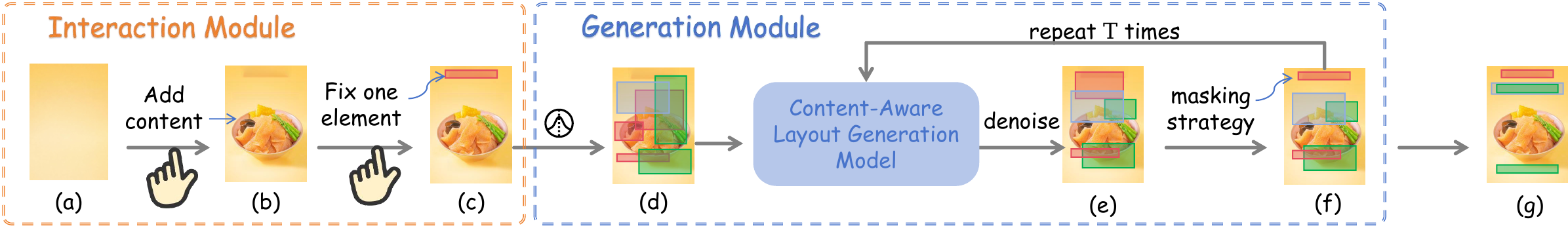}
\caption{
iPoster’s interactive framework, showing user interaction and mask execution during inference.
}
\label{fig:cond_pipe}
\end{figure*}

\subsection{Training and Inference Mechanisms}
\label{sec:training and inference}

\textbf{Training}. 
As shown in Figure~\ref{fig:pipe}, during the training process, GNNs are used as backbone networks to capture the high-dimensional spatial relationship between layout elements and the canvas.
The inputs from Section~\ref{sec:Model Inputs} are encoded according to the encoders of each branch to obtain three features: $F_{Bbox}$, $F_{Layout}$, and $F_{Image}$. Then, we construct two graphs according to Section~\ref{sec:Constructing Graphs}, $G_{BLM}$ and $G_{ILM}$. These are processed by the Cross Content-aware Attention Module to obtain content-balanced features. Finally, a noise prediction network predicts the noise $\epsilon_\theta$, and the loss is calculated using the ground truth noise $\epsilon$ and the predicted noise $\epsilon_\theta$.

\textbf{Interaction and Inference}. We regard the layout generation problem as a denoising diffusion process.
During inference, the user first selects a clean, content-free background image (see Figure~\ref{fig:cond_pipe}-$a$), and then specifies and positions the main elements of the poster (see Figure~\ref{fig:cond_pipe}-$b$). The user may impose any of the constraints described in Section~\ref{sec:Preliminaries} as dictated by the task requirements; for example, in the task $Completion$, specific layout elements can be fixed to preserve their spatial arrangement and structural attributes (see Figure~\ref{fig:cond_pipe}-$c$). The constrained layout and canvas are then provided to each branch for feature encoding and subsequent noise prediction. 
At each iteration of denoising, the model produces an intermediate denoising estimate (Figure~\ref{fig:cond_pipe}-$e$) and updates the layout using the masking strategy associated with the active constraint (Figure~\ref{fig:cond_pipe}-$f$), guiding subsequent denoising. 
After $T$ denoising iterations, the procedure yields a high-quality layout that conforms to the global layout distribution while satisfying user-specified constraints (Figure~\ref{fig:cond_pipe}-$g$).

\textbf{Masking Strategies}.
To enable a unified model to support diverse user-driven layout tasks, we employ \textit{task-adaptive masking strategies} during both training and inference. These masking strategies serve as the primary mechanism for injecting user-provided constraints into the generative process without requiring task-specific model modifications.

Concretely, at each denoising step $t$, the model predicts a layout representation $\mathbf{x}_t \in \mathbb{R}^{N \times 5}$, where each of the $N$ elements encodes a bounding box in the format $[c, c_x, c_y, w, h]$ (category, center coordinates, width, height). A binary mask $\mathbf{M} \in \{0,1\}^{N \times 5}$ is then applied to selectively preserve user-specified attributes while allowing the model to freely generate unconstrained dimensions. Specifically:
\begin{itemize}
    \item For $C \rightarrow S + P$, only the category dimension ($c$) is fixed via masking (i.e., $M_{i,0} = 1$), while spatial attributes ($c_x, c_y, w, h$) are predicted ($M_{i,j} = 0$ for $j=1,\dots,4$).
    \item For $C + S \rightarrow P$, both category and size ($c, w, h$) are constrained ($M_{i,0}=M_{i,3}=M_{i,4}=1$), and only position ($c_x, c_y$) is predicted.
    \item For \textit{Completion}, user-provided anchor elements have all attributes fixed ($\mathbf{M}_{i,:} = \mathbf{1}$), effectively freezing them throughout denoising; unspecified elements are predicted.
    \item For \textit{Refinement}, the input layout is regarded as an intermediate state in the reverse diffusion process, enabling optimization through a few denoising steps without additional training.
\end{itemize}

During inference, user-specified values $\mathbf{x}_{\text{user}}$ are enforced by overwriting $\hat{\mathbf{x}}_t$ wherever $\mathbf{M} = 1$:
\begin{equation*}
    \mathbf{x}_t \leftarrow \mathbf{M} \odot \mathbf{x}_{\text{user}} + (1 - \mathbf{M}) \odot \hat{\mathbf{x}}_t,
\end{equation*}
This ensures that user intent is exactly preserved at every denoising step, while the diffusion process jointly optimizes the remaining degrees of freedom in a context-aware manner.

\section{Evaluation}

\subsection{Datasets and Experiment Setup}
To evaluate the effectiveness of our model, we conducted comprehensive evaluations on two public datasets: CGL \cite{cglgan} and PKU \cite{posterlayout}. 
We compare the performance of several representative baseline methods, including 
CGL-GAN~\cite{cglgan}, 
RALF~\cite{ralf}, 
LayoutDiT~\cite{layoutdit}.
More details about the datasets and experimental setup can be found in the supplementary materials.

\subsection{Evaluation Metrics}
Referring to the previous models \cite{cglgan,posterlayout, ralf, layoutdit}, we selected the following two categories of metrics to evaluate the effectiveness of our model.

\textbf{Content Metrics.}
These metrics evaluate the harmony between the layouts and the canvas. 
Occlusion ($Occ$) measures the average overlap between layout elements and salient regions. 
Readability Scores ($Rea$) evaluates text region flatness by computing the average pixel gradient within text areas in the image space.

\textbf{Graphic Metrics.}
These metrics evaluate the quality of the generated layouts without considering the canvas.
Loose Underlay Validness ($Und_l$) evaluates the overlap ratio between the underlay and non-underlay elements.
Strict Underlay Validness ($Und_s$) calculates the effectiveness ratio of underlay elements. An underlay element is scored 1 only when it entirely covers a non-underlay element, otherwise, it scores 0.
Overlay ($Ove$) represents the average Intersection over Union (IoU) of all pairs of elements that do not contain the underlay elements.

\begin{figure*}[htb]
    \begin{minipage}{0.3\linewidth}
        \centering
        \includegraphics[width=0.9\linewidth]{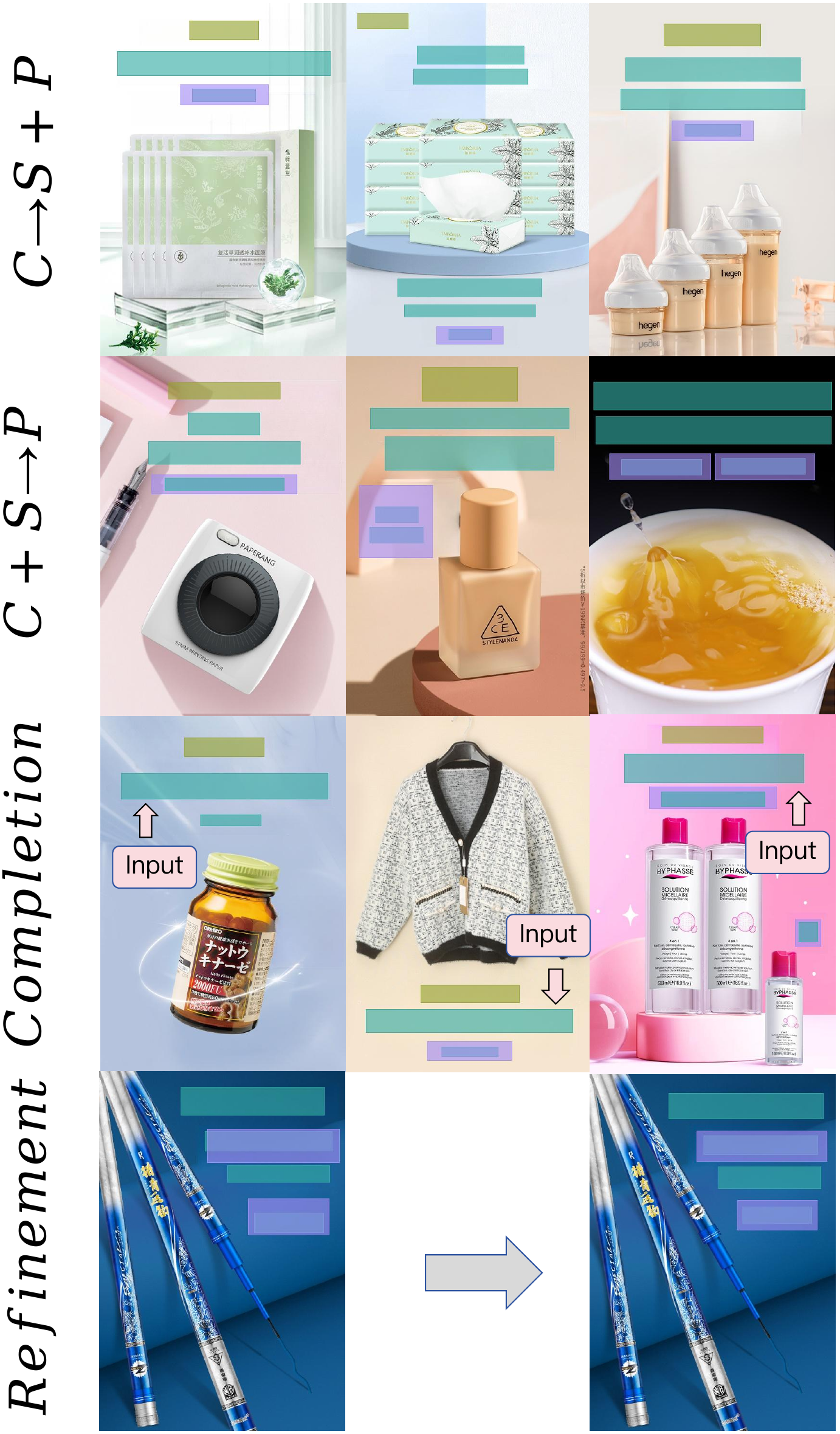}
        \captionof{figure}{
        Test examples of different constrained generation tasks.
        }
        \label{fig:cond}
    \end{minipage}
    \hfill
    \begin{minipage}{0.68\linewidth}
        \captionof{table}{Quantitative results of the different constrained generation tasks on the PKU and CGL annotated datasets.
        }
        \resizebox{\linewidth}{!}{
            \centering
            \begin{tabular}{lcccccccccc}
            \bottomrule
            \multirow{3}{*}{\textbf{Method}} & 
            \multicolumn{5}{c}{\textbf{PKU}} & \multicolumn{5}{c}{\textbf{CGL}} 
            \\
            \cmidrule(lr){2-6} \cmidrule(lr){7-11}
            & \multicolumn{2}{c}{\textbf{Content}} & \multicolumn{3}{c}{\textbf{Graphic}} 
            & \multicolumn{2}{c}{\textbf{Content}} & \multicolumn{3}{c}{\textbf{Graphic}}
            \\
            \cmidrule(lr){2-3} \cmidrule(lr){4-6} \cmidrule(lr){7-8} \cmidrule(lr){9-11} 
            &\(\text{$Occ$}\downarrow\) & \(\text{$Rea$}\downarrow\) & \(\text{$Und_l$}\uparrow\) & \(\text{$Und_s$}\uparrow\) & \(\text{$Ove$}\downarrow\) 
            &\(\text{$Occ$}\downarrow\) & \(\text{$Rea$}\downarrow\) & \(\text{$Und_l$}\uparrow\) & \(\text{$Und_s$}\uparrow\) & \(\text{$Ove$}\downarrow\) 
            \\ 
            \bottomrule
            
            \textbf{C} $\rightarrow$ \textbf{S}  + \textbf{P}
            \\    
            CGL-GAN     
                & 0.150         & 0.0173        & 0.690         & 0.480         & 0.0368       
                & 0.139         & 0.0217        & 0.891         & 0.608         & 0.0501 \\
            RALF        
                & \underline{0.126}         & \underline{0.0138}        & 0.968         & \underline{0.886}         & 0.0095       
                & 0.130         & 0.0182        & 0.986         &\textbf{0.964} & \underline{0.0064} \\
            LayoutDiT   
                & 0.148         & 0.0145        & \underline{0.978}         & 0.872         & \underline{0.0029}       
                &\textbf{0.120} & \textbf{0.0140}        & \underline{0.988}         & 0.937         & 0.0068 \\
            \textbf{iPoster(Ours)} 
                & \textbf{0.120} & \textbf{0.0133} & \textbf{0.994} & \textbf{0.978} & \textbf{0.0018} 
                & \underline{0.124} & \underline{0.0142} & \textbf{0.991} & \underline{0.956}     & \textbf{0.0034} 
            \\
            \bottomrule
            \textbf{C} $+$ \textbf{S} $\rightarrow$ \textbf{P}
            \\    
            CGL-GAN     
                & 0.135         & 0.0160        & 0.669         & 0.450         & 0.0440       
                & 0.141         & 0.0224        & 0.880         & 0.545         & 0.0459 \\
            RALF        
                & \underline{0.134}         & 0.0141        & 0.953         & 0.874         & 0.0103       
                & 0.130         & 0.0184        & \underline{0.985}         &\textbf{0.957} & \underline{0.0057} \\
            LayoutDiT   
                & 0.144         & \underline{0.0140}        & \underline{0.974}         & \underline{0.886}         & \underline{0.0082}      
                &\textbf{0.126} & \underline{0.0138}        & 0.969         & 0.843         & 0.0108 \\
            \textbf{iPoster(Ours)} 
                & \textbf{0.126} & \textbf{0.0134} & \textbf{0.984} & \textbf{0.911} & \textbf{0.0070} 
                & \underline{0.128} & \textbf{0.0132} & \textbf{0.989} & \underline{0.903}          & \textbf{0.0048} 
            \\
            \bottomrule
            \textbf{Completion}
            \\    
            CGL-GAN     
                & 0.153         & 0.0175        & 0.650         & 0.436         & 0.0633       
                & 0.199         & 0.0230        & 0.670         & 0.230         & 0.1504 \\
            RALF        
                & \underline{0.128}         &\textbf{0.0137}&\underline{0.960} &\textbf{0.883} & 0.0123       
                & 0.128         & 0.0183        &\underline{0.987}         &\textbf{0.963}         & 0.0050 \\
            LayoutDiT   
                & 0.130         & 0.0148        & 0.931         & 0.873         & \underline{0.0054}       
                &\textbf{0.119} & \underline{0.0149}        & 0.970         & 0.901         & \underline{0.0036} \\
            \textbf{iPoster(Ours)} 
                & \textbf{0.125} & \underline{0.0142}          & \textbf{0.968}      &\underline{0.881} & \textbf{0.0018} 
                & \underline{0.122}          & \textbf{0.0142} & \textbf{0.988}      & \underline{0.926}         & \textbf{0.0017} 
            \\
            \bottomrule
            \textbf{Refinement}
            \\    
            CGL-GAN     
                & 0.125         & 0.0146        & 0.680         & 0.403         & 0.0883       
                & 0.129         & 0.0190        & 0.900         & 0.811         & 0.0255 \\
            RALF        
                &\textbf{0.118} & 0.0110        & 0.988         & 0.950         & 0.0046       
                &\textbf{0.126} & 0.0176        &\underline{0.993} &\textbf{0.982} & 0.0025 \\
            LayoutDiT   
                & 0.128         & \underline{0.0104}        & \underline{0.990}         & \underline{0.966}         &\underline{0.0012}        
                & 0.141         & \underline{0.0113}        & 0.990         & 0.955         & \underline{0.0015} \\
            \textbf{iPoster(Ours)} 
                & \underline{0.125}         & \textbf{0.0101} & \textbf{0.992} &\textbf{0.970}& \textbf{0.0011} 
                & \underline{0.140}         & \textbf{0.0112} & \textbf{0.994} &\underline{0.972}       & \textbf{0.0013} 
            \\
            \bottomrule

            \end{tabular}
        }
        \label{tab:cond}
    \end{minipage}
\end{figure*}

\subsection{Constrained Generation}

Constrained experiments enable the customization of generated outputs to accommodate diverse user requirements, offering practical relevance in real-world applications. 
Based on the classification in Section~\ref{sec:Preliminaries}, we designed comprehensive test sets for four types of user constraint tasks and conducted experiments and analyses.
As the compared methods lack real-time interaction, we standardized the test set via preprocessing and integrated these baselines into our framework to ensure fair quantitative evaluation.

Figure~\ref{fig:cond} and Table~\ref{tab:cond}  show the qualitative and quantitative results of our model under different constrained tasks. 
For instance, iPoster achieves superior performance across all tasks in terms of the $Ove$ metric. Furthermore, the observed reduction in IoU between layout elements indicates a substantial mitigation of element overlapping.
These scenarios reflect common user interactions in real-world design workflows.
The results demonstrate that iPoster not only generates high-quality layouts but, more importantly, faithfully respects diverse user-provided constraints. 
This tight coupling between user intent and model behavior enables a responsive and predictable interactive experience, where users retain full control over critical elements while delegating low-level spatial reasoning to the system. 
More experimental results and poster samples can be found in the supplementary materials.

\begin{table}[htbp]
\vspace{-0.2cm}
\caption{
Comparison of model parameter counts and end-to-end inference latency for batch size of 1.
}
\centering
\resizebox{0.98\columnwidth}{!}{
    \begin{tabular}{lcccc}
        \hline
        \quad 
        &\(\text{PosterLlama}~\cite{seol2024posterllama}\) 
        & \(\text{PosterO}~\cite{hsu2025postero}\) 
        & \(\text{LayoutDiT}~\cite{layoutdit}\) 
        &\(\text{iPoster}\) 
    
        \\ 
        \midrule
    
        Params
            & 7B    & 8B     & 49M   & 33M  
        \\
        Inference Time(s) 
            & 7.6  & 7.5    & 1.5   & 1.1    
        \\
        \bottomrule
    
    \end{tabular}
}

\label{tab:time}

\vspace{-0.2cm}
\end{table}
We have also collected inference time statistics for several models in the entire processing pipeline, including LLM-based models and lightweight models, as shown in Table~\ref{tab:time}. All inference times were measured on a single NVIDIA A100 GPU (40 GB memory). For example, PosterO~\cite{hsu2025postero}, which employs Llama-3.1-8B as the backbone, requires approximately 7.5 seconds per image for inference. Among lightweight approaches. 
LayoutDiT achieves a per-image inference time of about 1.5 seconds with 49M parameters.
And our proposed iPoster achieves approximately 1.1 seconds with 33M parameters. 
It is sufficiently efficient to support real-time interaction, especially in comparison to LLM-based methods.
Furthermore, for iPoster, since different constrained generation tasks use a unified architecture with only different masking strategies, the inference time is basically the same for different tasks.

\subsection{User Scenario}
\begin{figure*}[htbp]
\centering
\includegraphics[width=1.\linewidth]{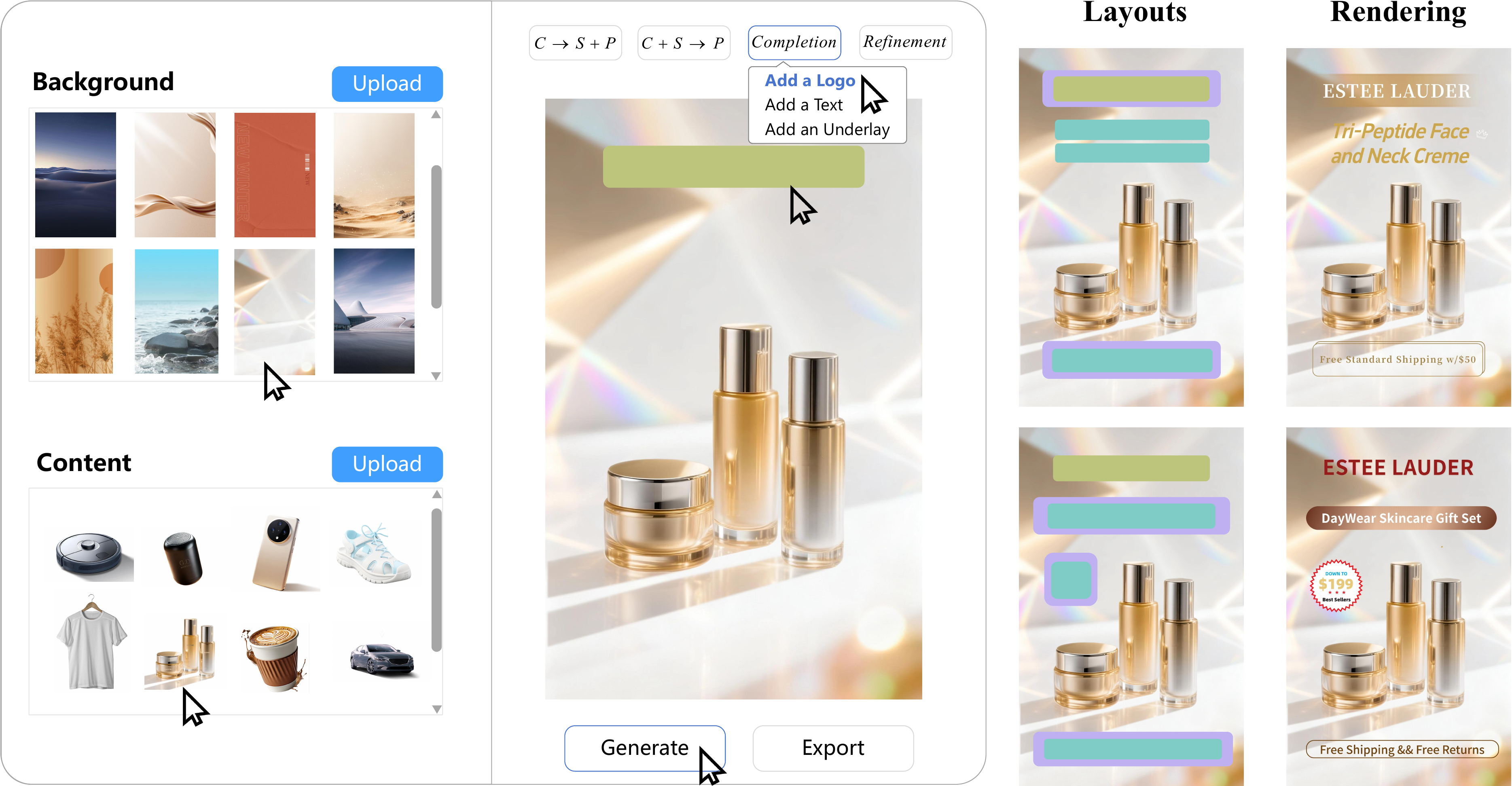}
\caption{
Illustration of a representative user scenario. The left side presents a schematic UI mockup depicting user interactions, and the right side shows the resulting layout candidates along with the final posters automatically rendered from them.
}

\label{fig:user_scenario}
\end{figure*}

Figure~\ref{fig:user_scenario} illustrates a prototype interface of a user scenario example.
The process begins with the ingestion of background and content materials through the left interface. Users then articulate the intent of the design applying constraints via the right-hand control panel; for example, selecting the task $Completion$ to add a logo to a suitable position on the poster. 
Following the specification of the desired sample count, the system generates a set of candidate layouts. Subsequently, these structural outputs guide an automated rendering engine, which precisely composites text and graphic into the final poster compositions.

\section{Discussion and Conclusion}
iPoster is an interactive, content-aware layout generation framework that combines graph-based spatial inference (via GNNs) with diffusion models and a Cross Content-aware Attention Module 
It dynamically aligns layout elements with image content, and incorporates flexible user constraints through masking.
Extensive experiments demonstrate robustness and state-of-the-art performance, offering a strong foundation for content-driven graphic design systems. 
Its user-friendly constraint interaction allows even non-professionals to quickly design product posters. Furthermore, its unique masking strategy enables various constraint tasks to operate within a unified framework.
However, it currently operates only at the visual hierarchy level and does not capture higher-level design semantics.
Future work will explore hierarchical graph representations, end-to-end rendering, and richer interaction modalities including sketches and language guidance.

\bibliographystyle{ACM-Reference-Format}
\bibliography{main}


\begin{thebibliography}{24}


\ifx \showCODEN    \undefined \def \showCODEN     #1{\unskip}     \fi
\ifx \showISBNx    \undefined \def \showISBNx     #1{\unskip}     \fi
\ifx \showISBNxiii \undefined \def \showISBNxiii  #1{\unskip}     \fi
\ifx \showISSN     \undefined \def \showISSN      #1{\unskip}     \fi
\ifx \showLCCN     \undefined \def \showLCCN      #1{\unskip}     \fi
\ifx \shownote     \undefined \def \shownote      #1{#1}          \fi
\ifx \showarticletitle \undefined \def \showarticletitle #1{#1}   \fi
\ifx \showURL      \undefined \def \showURL       {\relax}        \fi
\providecommand\bibfield[2]{#2}
\providecommand\bibinfo[2]{#2}
\providecommand\natexlab[1]{#1}
\providecommand\showeprint[2][]{arXiv:#2}

\bibitem[Cao et~al\mbox{.}(2022)]%
        {icvt}
\bibfield{author}{\bibinfo{person}{Yunning Cao}, \bibinfo{person}{Ye Ma}, \bibinfo{person}{Min Zhou}, \bibinfo{person}{Chuanbin Liu}, \bibinfo{person}{Hongtao Xie}, \bibinfo{person}{Tiezheng Ge}, {and} \bibinfo{person}{Yuning Jiang}.} \bibinfo{year}{2022}\natexlab{}.
\newblock \showarticletitle{Geometry aligned variational transformer for image-conditioned layout generation}. In \bibinfo{booktitle}{\emph{Proceedings of the ACM International Conference on Multimedia}}. \bibinfo{pages}{1561--1571}.
\newblock


\bibitem[Guo et~al\mbox{.}(2021)]%
        {guo2021vinci}
\bibfield{author}{\bibinfo{person}{Shunan Guo}, \bibinfo{person}{Zhuochen Jin}, \bibinfo{person}{Fuling Sun}, \bibinfo{person}{Jingwen Li}, \bibinfo{person}{Zhaorui Li}, \bibinfo{person}{Yang Shi}, {and} \bibinfo{person}{Nan Cao}.} \bibinfo{year}{2021}\natexlab{}.
\newblock \showarticletitle{Vinci: an intelligent graphic design system for generating advertising posters}. In \bibinfo{booktitle}{\emph{Proceedings of the CHI Conference on Human Factors in Computing Systems}}. \bibinfo{pages}{1--17}.
\newblock


\bibitem[Hegemann et~al\mbox{.}(2023)]%
        {hegemann2023computational}
\bibfield{author}{\bibinfo{person}{Lena Hegemann}, \bibinfo{person}{Yue Jiang}, \bibinfo{person}{Joon~Gi Shin}, \bibinfo{person}{Yi-Chi Liao}, \bibinfo{person}{Markku Laine}, {and} \bibinfo{person}{Antti Oulasvirta}.} \bibinfo{year}{2023}\natexlab{}.
\newblock \showarticletitle{Computational Assistance for User Interface Design: Smarter Generation and Evaluation of Design Ideas}. In \bibinfo{booktitle}{\emph{Proceedings of the CHI Conference on Human Factors in Computing Systems}}.
\newblock


\bibitem[Horita et~al\mbox{.}(2024)]%
        {ralf}
\bibfield{author}{\bibinfo{person}{Daichi Horita}, \bibinfo{person}{Naoto Inoue}, \bibinfo{person}{Kotaro Kikuchi}, \bibinfo{person}{Kota Yamaguchi}, {and} \bibinfo{person}{Kiyoharu Aizawa}.} \bibinfo{year}{2024}\natexlab{}.
\newblock \showarticletitle{Retrieval-augmented layout transformer for content-aware layout generation}. In \bibinfo{booktitle}{\emph{Proceedings of the IEEE/CVF Conference on Computer Vision and Pattern Recognition}}. \bibinfo{pages}{67--76}.
\newblock


\bibitem[Hsu and Peng(2025)]%
        {hsu2025postero}
\bibfield{author}{\bibinfo{person}{HsiaoYuan Hsu} {and} \bibinfo{person}{Yuxin Peng}.} \bibinfo{year}{2025}\natexlab{}.
\newblock \showarticletitle{PosterO: Structuring Layout Trees to Enable Language Models in Generalized Content-Aware Layout Generation}. In \bibinfo{booktitle}{\emph{Proceedings of the IEEE/CVF Conference on Computer Vision and Pattern Recognition}}. \bibinfo{pages}{8117--8127}.
\newblock


\bibitem[Hsu et~al\mbox{.}(2023)]%
        {posterlayout}
\bibfield{author}{\bibinfo{person}{Hsiao~Yuan Hsu}, \bibinfo{person}{Xiangteng He}, \bibinfo{person}{Yuxin Peng}, \bibinfo{person}{Hao Kong}, {and} \bibinfo{person}{Qing Zhang}.} \bibinfo{year}{2023}\natexlab{}.
\newblock \showarticletitle{Posterlayout: A new benchmark and approach for content-aware visual-textual presentation layout}. In \bibinfo{booktitle}{\emph{Proceedings of the IEEE/CVF Conference on Computer Vision and Pattern Recognition}}. \bibinfo{pages}{6018--6026}.
\newblock


\bibitem[Jahanian et~al\mbox{.}(2013)]%
        {jahanian2013recommendation}
\bibfield{author}{\bibinfo{person}{Ali Jahanian}, \bibinfo{person}{Jerry Liu}, \bibinfo{person}{Qian Lin}, \bibinfo{person}{Daniel Tretter}, \bibinfo{person}{Eamonn O'Brien-Strain}, \bibinfo{person}{Seungyon~Claire Lee}, \bibinfo{person}{Nic Lyons}, {and} \bibinfo{person}{Jan Allebach}.} \bibinfo{year}{2013}\natexlab{}.
\newblock \showarticletitle{Recommendation system for automatic design of magazine covers}. In \bibinfo{booktitle}{\emph{Proceedings of the International Conference on Intelligent User Interfaces}}. \bibinfo{pages}{95--106}.
\newblock


\bibitem[Jiang et~al\mbox{.}(2023)]%
        {jiang2023layoutformer}
\bibfield{author}{\bibinfo{person}{Zhaoyun Jiang}, \bibinfo{person}{Jiaqi Guo}, \bibinfo{person}{Shizhao Sun}, \bibinfo{person}{Huayu Deng}, \bibinfo{person}{Zhongkai Wu}, \bibinfo{person}{Vuksan Mijovic}, \bibinfo{person}{Zijiang~James Yang}, \bibinfo{person}{Jian-Guang Lou}, {and} \bibinfo{person}{Dongmei Zhang}.} \bibinfo{year}{2023}\natexlab{}.
\newblock \showarticletitle{Layoutformer++: Conditional graphic layout generation via constraint serialization and decoding space restriction}. In \bibinfo{booktitle}{\emph{Proceedings of the IEEE/CVF Conference on Computer Vision and Pattern Recognition}}. \bibinfo{pages}{18403--18412}.
\newblock


\bibitem[Kong et~al\mbox{.}(2022)]%
        {kong2022blt}
\bibfield{author}{\bibinfo{person}{Xiang Kong}, \bibinfo{person}{Lu Jiang}, \bibinfo{person}{Huiwen Chang}, \bibinfo{person}{Han Zhang}, \bibinfo{person}{Yuan Hao}, \bibinfo{person}{Haifeng Gong}, {and} \bibinfo{person}{Irfan Essa}.} \bibinfo{year}{2022}\natexlab{}.
\newblock \showarticletitle{Blt: Bidirectional layout transformer for controllable layout generation}. In \bibinfo{booktitle}{\emph{European Conference on Computer Vision}}. Springer, \bibinfo{pages}{474--490}.
\newblock


\bibitem[Lee et~al\mbox{.}(2010)]%
        {lee2010designing}
\bibfield{author}{\bibinfo{person}{Brian Lee}, \bibinfo{person}{Savil Srivastava}, \bibinfo{person}{Ranjitha Kumar}, \bibinfo{person}{Ronen Brafman}, {and} \bibinfo{person}{Scott~R Klemmer}.} \bibinfo{year}{2010}\natexlab{}.
\newblock \showarticletitle{Designing with interactive example galleries}. In \bibinfo{booktitle}{\emph{Proceedings of the CHI Conference on Human Factors in Computing Systems}}. \bibinfo{pages}{2257--2266}.
\newblock


\bibitem[Li et~al\mbox{.}(2023)]%
        {radm}
\bibfield{author}{\bibinfo{person}{Fengheng Li}, \bibinfo{person}{An Liu}, \bibinfo{person}{Wei Feng}, \bibinfo{person}{Honghe Zhu}, \bibinfo{person}{Yaoyu Li}, \bibinfo{person}{Zheng Zhang}, \bibinfo{person}{Jingjing Lv}, \bibinfo{person}{Xin Zhu}, \bibinfo{person}{Junjie Shen}, \bibinfo{person}{Zhangang Lin}, {et~al\mbox{.}}} \bibinfo{year}{2023}\natexlab{}.
\newblock \showarticletitle{Relation-aware diffusion model for controllable poster layout generation}. In \bibinfo{booktitle}{\emph{Proceedings of the ACM International Conference on Information and Knowledge Management}}. \bibinfo{pages}{1249--1258}.
\newblock


\bibitem[Li et~al\mbox{.}(2024)]%
        {layoutdit}
\bibfield{author}{\bibinfo{person}{Yu Li}, \bibinfo{person}{Yifan Chen}, \bibinfo{person}{Gongye Liu}, \bibinfo{person}{Fei Yin}, \bibinfo{person}{Qingyan Bai}, \bibinfo{person}{Jie Wu}, \bibinfo{person}{Hongfa Wang}, \bibinfo{person}{Ruihang Chu}, {and} \bibinfo{person}{Yujiu Yang}.} \bibinfo{year}{2024}\natexlab{}.
\newblock \showarticletitle{LayoutDiT: Exploring Content-Graphic Balance in Layout Generation with Diffusion Transformer}.
\newblock \bibinfo{journal}{\emph{arXiv preprint arXiv:2407.15233}} (\bibinfo{year}{2024}).
\newblock


\bibitem[Lin et~al\mbox{.}(2023)]%
        {lin2023layoutprompter}
\bibfield{author}{\bibinfo{person}{Jiawei Lin}, \bibinfo{person}{Jiaqi Guo}, \bibinfo{person}{Shizhao Sun}, \bibinfo{person}{Zijiang Yang}, \bibinfo{person}{Jian-Guang Lou}, {and} \bibinfo{person}{Dongmei Zhang}.} \bibinfo{year}{2023}\natexlab{}.
\newblock \showarticletitle{Layoutprompter: Awaken the design ability of large language models}.
\newblock \bibinfo{journal}{\emph{Advances in Neural Information Processing Systems}}  \bibinfo{volume}{36} (\bibinfo{year}{2023}), \bibinfo{pages}{43852--43879}.
\newblock


\bibitem[O'Donovan et~al\mbox{.}({[n.\,d.]})]%
        {o2015designscape}
\bibfield{author}{\bibinfo{person}{Peter O'Donovan}, \bibinfo{person}{Aseem Agarwala}, {and} \bibinfo{person}{Aaron Hertzmann}.} \bibinfo{year}{[n.\,d.]}\natexlab{}.
\newblock \showarticletitle{Designscape: Design with interactive layout suggestions}. In \bibinfo{booktitle}{\emph{Proceedings of the CHI Conference on Human Factors in Computing Systems}}. \bibinfo{pages}{1221--1224}.
\newblock


\bibitem[Patnaik et~al\mbox{.}(2025)]%
        {patnaik2025aesthetiq}
\bibfield{author}{\bibinfo{person}{Sohan Patnaik}, \bibinfo{person}{Rishabh Jain}, \bibinfo{person}{Balaji Krishnamurthy}, {and} \bibinfo{person}{Mausoom Sarkar}.} \bibinfo{year}{2025}\natexlab{}.
\newblock \showarticletitle{AesthetiQ: Enhancing Graphic Layout Design via Aesthetic-Aware Preference Alignment of Multi-modal Large Language Models}. In \bibinfo{booktitle}{\emph{Proceedings of the IEEE/CVF Conference on Computer Vision and Pattern Recognition}}. \bibinfo{pages}{23701--23711}.
\newblock


\bibitem[Scarselli et~al\mbox{.}(2008)]%
        {scarselli2008graph}
\bibfield{author}{\bibinfo{person}{Franco Scarselli}, \bibinfo{person}{Marco Gori}, \bibinfo{person}{Ah~Chung Tsoi}, \bibinfo{person}{Markus Hagenbuchner}, {and} \bibinfo{person}{Gabriele Monfardini}.} \bibinfo{year}{2008}\natexlab{}.
\newblock \showarticletitle{The graph neural network model}.
\newblock \bibinfo{journal}{\emph{IEEE Transactions on Neural Networks}} \bibinfo{volume}{20}, \bibinfo{number}{1} (\bibinfo{year}{2008}), \bibinfo{pages}{61--80}.
\newblock


\bibitem[Seol et~al\mbox{.}(2024)]%
        {seol2024posterllama}
\bibfield{author}{\bibinfo{person}{Jaejung Seol}, \bibinfo{person}{Seojun Kim}, {and} \bibinfo{person}{Jaejun Yoo}.} \bibinfo{year}{2024}\natexlab{}.
\newblock \showarticletitle{Posterllama: Bridging design ability of language model to content-aware layout generation}. In \bibinfo{booktitle}{\emph{Proceedings of the Conference on European Conference on Computer Vision}}. Springer, \bibinfo{pages}{451--468}.
\newblock


\bibitem[Shi et~al\mbox{.}(2025)]%
        {shi2025layoutcot}
\bibfield{author}{\bibinfo{person}{Hengyu Shi}, \bibinfo{person}{Junhao Su}, \bibinfo{person}{Huansheng Ning}, \bibinfo{person}{Xiaoming Wei}, {and} \bibinfo{person}{Jialin Gao}.} \bibinfo{year}{2025}\natexlab{}.
\newblock \showarticletitle{LayoutCoT: Unleashing the Deep Reasoning Potential of Large Language Models for Layout Generation}.
\newblock \bibinfo{journal}{\emph{arXiv preprint arXiv:2504.10829}} (\bibinfo{year}{2025}).
\newblock


\bibitem[Shi et~al\mbox{.}(2020)]%
        {shi2020emog}
\bibfield{author}{\bibinfo{person}{Yang Shi}, \bibinfo{person}{Nan Cao}, \bibinfo{person}{Xiaojuan Ma}, \bibinfo{person}{Siji Chen}, {and} \bibinfo{person}{Pei Liu}.} \bibinfo{year}{2020}\natexlab{}.
\newblock \showarticletitle{EmoG: supporting the sketching of emotional expressions for storyboarding}. In \bibinfo{booktitle}{\emph{Proceedings of the CHI Conference on Human Factors in Computing Systems}}. \bibinfo{pages}{1--12}.
\newblock


\bibitem[Yang et~al\mbox{.}(2024)]%
        {posterllava}
\bibfield{author}{\bibinfo{person}{Tao Yang}, \bibinfo{person}{Yingmin Luo}, \bibinfo{person}{Zhongang Qi}, \bibinfo{person}{Yang Wu}, \bibinfo{person}{Ying Shan}, {and} \bibinfo{person}{Chang~Wen Chen}.} \bibinfo{year}{2024}\natexlab{}.
\newblock \showarticletitle{Posterllava: Constructing a unified multi-modal layout generator with llm}.
\newblock \bibinfo{journal}{\emph{arXiv preprint arXiv:2406.02884}} (\bibinfo{year}{2024}).
\newblock


\bibitem[Yang et~al\mbox{.}(2016)]%
        {yang2016automatic}
\bibfield{author}{\bibinfo{person}{Xuyong Yang}, \bibinfo{person}{Tao Mei}, \bibinfo{person}{Ying-Qing Xu}, \bibinfo{person}{Yong Rui}, {and} \bibinfo{person}{Shipeng Li}.} \bibinfo{year}{2016}\natexlab{}.
\newblock \showarticletitle{Automatic generation of visual-textual presentation layout}.
\newblock \bibinfo{journal}{\emph{ACM Transactions on Multimedia Computing, Communications, and Applications (TOMM)}} \bibinfo{volume}{12}, \bibinfo{number}{2} (\bibinfo{year}{2016}), \bibinfo{pages}{1--22}.
\newblock


\bibitem[Zhao et~al\mbox{.}(2020)]%
        {zhao2020iconate}
\bibfield{author}{\bibinfo{person}{Nanxuan Zhao}, \bibinfo{person}{Nam~Wook Kim}, \bibinfo{person}{Laura~Mariah Herman}, \bibinfo{person}{Hanspeter Pfister}, \bibinfo{person}{Rynson~WH Lau}, \bibinfo{person}{Jose Echevarria}, {and} \bibinfo{person}{Zoya Bylinskii}.} \bibinfo{year}{2020}\natexlab{}.
\newblock \showarticletitle{Iconate: Automatic compound icon generation and ideation}. In \bibinfo{booktitle}{\emph{Proceedings of the CHI Conference on Human Factors in Computing Systems}}.
\newblock


\bibitem[Zheng et~al\mbox{.}(2019)]%
        {zheng2019content}
\bibfield{author}{\bibinfo{person}{Xinru Zheng}, \bibinfo{person}{Xiaotian Qiao}, \bibinfo{person}{Ying Cao}, {and} \bibinfo{person}{Rynson~WH Lau}.} \bibinfo{year}{2019}\natexlab{}.
\newblock \showarticletitle{Content-aware generative modeling of graphic design layouts}.
\newblock \bibinfo{journal}{\emph{ACM Transactions on Graphics (TOG)}} \bibinfo{volume}{38}, \bibinfo{number}{4} (\bibinfo{year}{2019}), \bibinfo{pages}{1--15}.
\newblock


\bibitem[Zhou et~al\mbox{.}(2022)]%
        {cglgan}
\bibfield{author}{\bibinfo{person}{Min Zhou}, \bibinfo{person}{Chenchen Xu}, \bibinfo{person}{Ye Ma}, \bibinfo{person}{Tiezheng Ge}, \bibinfo{person}{Yuning Jiang}, {and} \bibinfo{person}{Weiwei Xu}.} \bibinfo{year}{2022}\natexlab{}.
\newblock \showarticletitle{Composition-aware Graphic Layout GAN for Visual-Textual Presentation Designs}. In \bibinfo{booktitle}{\emph{Proceedings of the International Joint Conference on Artificial Intelligence}}. International Joint Conferences on Artificial Intelligence Organization, \bibinfo{pages}{4995--5001}.
\newblock


\end{thebibliography}


\end{document}